\documentclass[aps,prl,reprint]{revtex4-1}  
\usepackage{graphicx}  
\usepackage{dcolumn}   
\usepackage{bm}        
\usepackage{subfigure}
\usepackage{draftcopy}
\begin{document}


\title{ Cooling, conductance and thermometric  performance of non-ideal normal metal-superconductor tunnel junction pairs}
\author{S. Chaudhuri} 
\author{I. J. Maasilta}
 \affiliation{Nanoscience Center, Department of Physics, P. O. Box 35, FIN-40014 University of Jyv\"askyl\"a, Finland}
 \email{maasilta@jyu.fi }


\begin{abstract}
We have investigated the effect of a difference in the  tunnelling resistances  of the individual normal metal-insulator-superconductor (NIS) tunnel junctions in a double junction SINIS device, with particular emphasis on the impact on the conductance, cooling and thermometric performance.   By  solving the electrical and thermal equations of the junctions in a self-consistent way,  we find that asymmetry gives  rise to many new features, such as appearance of an excess sub-gap current, improved cooling performance,  exhibition of negative differential resistance, and improved temperature range of thermometric sensitivity.  Experiments were also carried out to  complement some of the numerical  results.  In addition, we studied theoretically and experimentally the effect of a finite series resistance, which also causes an excess current in the subgap region, and a suppression of the conductance maxima at the gap edge.  Experimental results agree well with the theoretical predictions. 
\end{abstract}

\maketitle

\section{Introduction}
For some time now normal metal-insulator-superconductor (NIS) tunnel junctions devices have been used as  solid state coolers and accurate low temperature thermometers in the sub-Kelvin temperature range \cite{nahum2, Nahum, RMP}. The concept of NIS solid state cooling  is based on the selective extraction of hot electrons from  the Fermi tail of the normal metal, using the finite energy gap of the superconductor as a filter, thereby reducing the average energy of the normal metal electrons. Thermometry with NIS junctions \cite{rowell,JLTP} has been used in the study of photonic \cite{pekolanature}, phononic \cite{yung,panu}, electron-phonon \cite{jenni,jenniJLTP,timofeev,underwood} and electronic \cite{peltonen} heat flow, and applications in bolometry have been envisioned \cite{nahum2,schmidt,kuzmin}. Cooling has been demostrated for both electrons \cite{Nahum,leivo,oneill} and phonons \cite{courtois,panu}, with high power ( $\sim$ 50 pW) coolers being able to cool external payloads from 300 mK to 190 mK \cite{nist}. In addition, NIS junctions have  recently been ulitized as accurate charge pumps for metrological applications \cite{natphys}.  In most of these applications, it is beneficial to connect two junctions in series to form a SINIS device, essentially doubling the performance. 

Traditionally, the performance of SINIS devices has been analyzed with two simplifying assumptions -  (i) the device is symmetric, i.e., the junction resistances are equal, and (ii) the normal metal has a vanishingly small resistance compared with the tunneling resistances. While these assumptions are fair for many devices, one should, nevertheless,  study the effects of relaxing the above assumptions, especially when considering novel materials and more complex device processing steps, where control over device parameters is more difficult. 

In this paper, we present theoretical  and some corresponding  experimental results on the impact of asymmetric tunnelling resistances, and of a finite normal metal series  resistance on the conductance, cooling, and thermometric  characteristics of  SINIS  double tunnel junction devices. The evolution of the characteristics as a function of increasing asymmetry and series resistance   has been investigated.  A knowledge of these results is not only essential for a more accurate interpretation of  SINIS junction based cooling and thermometric experiments, but they can also  be exploited to optimize the performance of such devices. 

\section{system under study}
 
   A schematic of a SINIS device is shown in Figure 1, where the two junctions are denoted as the left  ($ L $) and the right ($ R $)  junction for convenience, with tunnelling resistances  $ R_{L} $ and $ R_{R} $, respectively. The  resistance of the  normal metal island is $ R_{N} $.  In the case of two probe measurements, one also has to consider the effect  of the probe resistances ($ R_{P} $), such as the line and the contact resistances. These appear in series with the junction resistances, and  can be lumped together with $ R_{N} $ into one series resistance $ R_{S}= R_{N}+R_{P}$  in the electrical circuit equations. The total applied voltage $V_{T}$ across the device is thus 
\begin{equation} \label{s3}  
  V_{T}=V_{L}+V_{R}+V_{S}.
\end{equation}  
   In four probe measurements $ V_{S} $  and $ R_{S} $  are replaced by  $ V_{N} $  and $ R_{N} $,  respectively.
 On the other hand, the finite resistance of the normal metal island of the device $ R_{N} $ means that there is power dissipation in the normal metal in the form of Joule heating. This inadvertent heating of the device can have an impact  on the conductance, cooling as well as thermometry aspects through the coupling of thermal and electrical circuit equations.  The Joule heating  associated with the probe resistances $ R_{P} $, however, does not affect the characteristics of the device, as long as all that power is absorbed by the refirgerator far from the device. We therefore ignore it in this study.

  \begin{figure}
\includegraphics[width=0.35\textwidth]{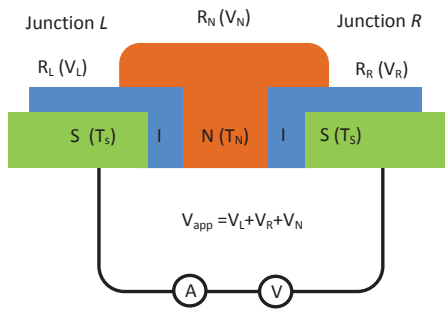}
\caption{[Color online] A Schematic of a SINIS device consisting of two NIS tunnel  junction connected in series having tunnelling resistances $ R_{L} $ and $ R_{R} $. $ R_{N} $  is the resistance of the normal metal. $ T_{S} $  and $ T_{N} $  are the electron temperatures of the superconductor and the normal metal respectively. } \label{SINIS}
\end{figure}
 Traditionally, an ideal SINIS junction is analyzed under the following  assumptions - (i) the tunnelling resistances of the individual junctions are equal  ($R_{L}$ = $ R_{R} $)  and (ii) the normal metal resistance is negligibly small   ($ R_{N}  \ll  R_{L}, R_{R} $).  This means that the total voltage drop across,  and the cooling power of an ideal  symmetric  SINIS device is  simply twice of  that for the corresponding individual NIS  junction: $ V_{T}=2V_{L}=2V_{R} $ and   $ R_{tot}=2R_{L}=2R_{R} $. Experimentally, however, these assumptions are not strictly true as inaccuracies in lithography and variations in local film thickness and morphology can give rise to a difference in the tunnelling resistances of the two junctions, even when fabricated  otherwise under the same conditions. Moreover, in most actual real devices, there is no extra Ohmic contact to the normal metal, and therefore a direct determination of the different resistance components is impossible.  This means that only the total resistance ($ R_{tot}=R_{L}+R_{R}+R_{S} $) of the SINIS device is determined form the high bias ($ V_{app} \gg 2\Delta $) slope of the current $I$ vs applied voltage $V_{app}$ curve,  and the possible effects of asymmetry and series resistances are not easily determined.    
 
   For these reasons, it is important to understand how the characteristics of a SINIS device evolve as a function of the uncertainty associated with the precise  determination of  these individual resistances.  Pekola \textit{et. al.}  \cite{Pekola} compared theoretically the cooling power of a couple of different asymmetric structures at a particular temperature,  and showed that the maximum  cooling power decreased only a little with increasing asymmetry. However, their study was not exhaustive, and a regime where cooling can be {\em enhanced} can be found, as we show in this paper. Also, it is possible that the value of $ R_{N} $ varies independently of the tunnelling resistances,  which  experimentally corresponds to changing the material of the normal metal island but keeping all other device  parameters the same, for example. 
For the sake of simplicity,  we consider here the following two distinct cases:\\
(i)\textit{Effect of asymmetric tunnelling resistance: $ R_{L}  \neq R_{R} $ and $ R_{S} $=0 :}
If the individual junction tunneling resistances $ R_{L} $ and $ R_{R} $ differ,  the corresponding voltages across the individual junctions ($ V_{L} $ and $ V_{R} $ )  are no longer identical although  the same current flows through both of them.  Similarly, the cooling power of the individual junctions may vary significantly.  \\
(ii)\textit{Effect of finite series resistance: $ R_{L}  = R_{R} $ and $ R_{S} $ $ \neq $0 :}
 The second goal is  to understand the impact of the finite series resistances, either in the device itself or external.  The effects are studied in a symmetric device, but could easily be  extended to asymmetric cases.

\section{Methods}
\subsubsection{Theory}

In this work we only consider first-order, single particle tunneling theory \cite{tink,Jochum}, and disregard higher order two-particle processes \cite{hekk1,Hekking,pekolaaverin} for the clarity of discussion. The higher order terms become more important with transparent barriers (low tunneling resistances) and with high resistivity of the normal metal; in many cases the junctions are opaque enough and/or the resisitivity small enough for the effect to be small. The standard result for the current in one voltage biased NIS junction is then

\begin{equation} \label{eqn:master}
I_{NIS} =\frac{1}{eR_{i}}\int_{-\infty}^{\infty}\!\! d\epsilon N_{S}(\epsilon )(f_{S}(\epsilon )-f_{N}(\epsilon+eV ))
\end{equation}
  
%
where   $ \epsilon $ is the energy, $ R_{i}= \hbar/(4\pi e^{2} A \vert t_{0}\vert^{2} N_{0,N}N_{0,S})$ is the tunnelling resistance of junction $i$, determined by the area of the junction $A$, the tunneling probablility $\vert t_{0}\vert^{2}$ and the normal state Fermi-energy densities of states of each lead $N_{0,N/S}$.  $f_{i}(\epsilon)$ is the Fermi function in  electrode $ i $, and $N_{S}(\epsilon )$ is the normalized superconducting quasiparticle density of states (DOS).  Here, instead of using just the equation from BCS theory, we use the expression introduced by Dynes et al. \cite{Dynes} to take into account non-idealities:  
\begin{equation} \label{eqn:dossc}
 D_{S}(\epsilon, T_{S})  =\left | {\rm Re} \left ( \frac{\epsilon+i\Gamma}{\sqrt{(\epsilon+i\Gamma)^{2}-\Delta(T _{S} )^{2}}} \right ) \right |,
 \end{equation}
where $ \Gamma $  is the Dynes parameter describing the broadening of the DOS singularity, and $\Delta(T_{S})$  is the superconducting energy gap, with its temperature dependence written explicitly. This equation is especially appropriate in Al junctions, where it was shown \cite{pekoladynes} to result from environmentally assisted tunneling events (stray photon assisted tunneling). One should also point out that, fortuitously, it can also parametrize higher order tunneling surprisingly well, including the subgap heating \cite{panuthesis}.  
  
 Since the individual junctions are in series, the same electrical current flows through both of them, although the voltages $V_{L}$, $V_{R}$ differ. However, the heat current is not the same. The expressions for current $I_i$ and heat current  $\dot{Q}_i$ through each junction ($i=L,R$) are thus given by the following equations \cite{RMP}:

\begin{widetext}
\begin{eqnarray}\label{I}
I_{} (V_{R/L},T_{N},T_{S}) &=&\frac{1}{eR_{L}}\int_{-\infty}^{\infty}\!\! d\epsilon N_{S}(\epsilon,T_{s})[f_{S}(\epsilon, T_{S} )-f_{N}(\epsilon+eV_{L},T_{N} )]\nonumber \\
   &=& \frac{1}{eR_{R}}\int_{-\infty}^{\infty}\!\! d\epsilon N_{S}(\epsilon +eV_{R}, T_{S})[f_{N}(\epsilon, T_{N} )-f_{S}(\epsilon+eV_{R},T_{S} )], \nonumber\\
\end{eqnarray}
\begin{equation} \label{PL}
\dot{Q}_{L}(V_{L,}T_{N},T_{S})=\frac{1}{e^{2}R_{L}}\int_{-\infty}^{\infty}\!\!d\epsilon(\epsilon+eV_{L} )N_{S}(\epsilon,T_{S} )[f_{S}(\epsilon,T_{S} )-f_{N}(\epsilon+V_{L},T_{N} )],
\end{equation}
\begin{equation} \label{PR}
\dot{Q}_{R} (V_{R},T_{N},T_{S}) =\frac{1}{e^{2}R_{R}}\int_{-\infty}^{\infty}\!\!d\epsilon(\epsilon)N_{S}(\epsilon +eV_{R}, T_{S})[f_{N}(\epsilon, T_{N} )-f_{S}(\epsilon+eV_{R},T_{S} )].
\end{equation}
\end{widetext}

The total cooling power of a SINIS device is thus given by  
 \begin{equation} \label{eqn:sum}
P_{T}=   -{ \dot{Q}_{L}} + \dot{Q}_{R},
 \end{equation}
where the minus sign takes care of the fact that $ P_{T} $ describes the total heat power extracted from the normal metal.  The situation where  $ P_{T}  $  is positive (negative) thus  corresponds to cooling (heating) of the normal metal.   In dynamic equilibrium, this cooling (heating) by the junctions is balanced by the inflow (outflow) of heat from (to) the surroundings, which leads to the equation
\begin{equation} \label{th}
P_{T} = B(T_{Bath}^{n}-T_{N}^{n})+\beta \left[ (P_{T}+I(V_{L}+V_{R}) \right ]+ I^{2}R_{N},
\end{equation}
which describes three distinct heating mechanisms:  The first term on the right gives the direct heating from the substrate (phonons at temperature $T_{bath}$), which couples to the normal metal electrons via electron-phonon interaction \cite{Wellstood, jenniJLTP, jenni}. $ B $ is the coupling strength constant which in the present case is given by $ B $ = $ \Sigma\Omega $ where  $ \Sigma$  is the electron-phonon coupling constant and  $\Omega $ is the normal metal electron gas volume.  The exponent $ n $ = 4-6  depends on disorder level, phonon dimensionality and film thickness \cite{jenni, lasse, Sergeev, geller}, but is typically $n = 5$ for good metals with thickness $> 30$ nm. In suspended wires, however, phonon transmission out of the nanowire becomes the bottleneck for heat flow \cite{panu}, which leads to $B$ being a geometry-dependent constant and $n \approx 2.8$. The second term describes the indirect back-flow of heat from the superconducting electrodes, associated with, for example, the quasi-particle recombination and backtunneling in the superconductor \cite{Fisher, oneill}. Here (0 $ \leq  \beta \leq $1)  is the fraction of total heat deposited in the superconductor flowing back to the normal metal. Finally, the third term is simply the direct Joule heating of the normal metal. Throughout this paper, we take the simplifying assumption that the superconductor temperature equals the bath temperature, $ T_{S}=T_{Bath} $.  This limit is useful to consider, as it gives the maximum performance that a cooler could achieve. A real device can also suffer from overheating and/or non-equilibrium effects of the superconductor, which degrade the cooler performance \cite{vasenko, rajauria,oneill}.

In this work, we have carried out numerical calculations for two distinct models which we name (1) electrical and (2) electrical + thermal, for convenience.  For the first case the calculations are carried out by only considering  the current and voltage equations (\ref{s3}) and (\ref{I}). In this case, all temperatures are equal $ T_{Bath} $  = $ T_{S} $ = $ T_{N} $. In practice, this limit corresponds to the case where either the normal metal volume is large (large $B$), or the tunneling resistances very high (small $P_{T}$ and $I$).  In the second case, both the electrical [Eqs.(\ref{s3}) and (\ref{I}) and the thermal equations [ Eq. (\ref{th})] are solved  simultaneously in a self-consistent way. In the second case, it is then possible to determine the value of $ T_{N} $ for a given applied bias and  $ T_{Bath} $ , but it is still assumed that $ T_{S} $ =  $ T_{Bath} $.  Unless explicitly mentioned, the following typical parameter  values were used for the simulations : $ \Gamma/\Delta $ =2 $\times $ 10$^{-4}$, $ \beta $ =0.03 (from Ref. \cite{panu}), $ n $=5 and $ B'$ =200, where we have defined a dimensionless e-p coupling constant $B'=e^2\Delta^{n-2}R_{T}B/(1.764k_B)^n$ and $R_{T} = R_{L}+R_{R}$. For typical AlO$_{x}$ based tunnel junctions with a tunneling resistance of 10 k$ \Omega $, $ \Delta $ =220 $ \mu $eV (Al superconductor) and $ \Sigma $ = 1.8 $ \times $10$ ^{9} $ (Cu normal metal), the value of $ B' $ = 200 corresponds to a normal metal electron volume of roughly 0.09 $\mu $m$^{3} $ (for example dimensions = 10 $\mu$m$ \times $ 300 nm$ \times 30$ nm, the normal metal dimensions of our experimental sample discussed below).    Throughout this paper, for the simulation results, the voltage, current, power, resistance and temperature are expressed in units of $ \Delta $/$ e $,  $ \Delta /eR_{T}$, $\Delta^{2}/e^{2}R_{T}$, $ R_{T}$ and $ T_{C}$, respectively.

\subsubsection{Experimental}
\label{sample}
  \begin{figure}
\includegraphics[width=0.35\textwidth]{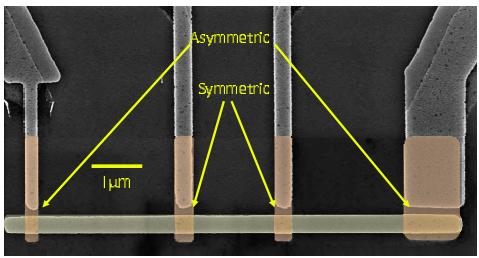}
\caption{[Color online] Scanning electron micrograph of a SINIS device with four NIS tunnel  junctions. Two SINIS pairs of junctions can be connected in such a way that both pairs have about the same total tunneling resistance, but one pair is asymmetric and the other symmetric. }   \label{sem}
\end{figure}

NIS junctions with Al, AlOx  and Cu  as the superconductor, insulator and the  normal metal, respectively, were fabricated on oxidized silicon substrates  using electron beam lithography and shadow evaporation techniques \cite{panu}. The thickness of Cu and Al were $\sim$ 30 and 20 nm, respectively. The insulator was formed by oxidation of the aluminium surface at room temperature at 400 mbar for 4 min in pure oxygen. In order to study the effect of asymmetry of the tunnelling resistances, we fabricated  SINIS devices  with the geometry depicted in Fig.$~\ref{sem}$. The asymmetric SINIS pair consists of  two  NIS junctions with junction areas 1030$\times$330 nm$^2$ and 210$\times$330  nm$^2$ (outer junctions), while the corresponding symmetric junction pair (inner junctions) have dimensions 350$\times$330 nm$^2$. The asymmetric SINIS had a total tunnelling resistance of 42.4 k$\Omega$, quite close to the value for the corresponding symmetric SINIS,  $\sim$ 44.4 k$\Omega$. The measurements were carried out using a He$^3$-He$^4$ dilution refrigerator with a base temperature of $\sim$  60 mK. Each measurement line had two   RC filters, one at 4 K and the other at 60 mK, contributing to a total resistance of $\sim$ 2 k$\Omega$ per line. In addition, additional microwave filtering was achieved between the RC filters with the help of Thermocoax cables \cite{thermocoax}.  DC I-V curves were measured using voltage and current preamplifiers (DL Instruments 1201 and 1211), and conductance was measured using a lock-in measurement with an excitation voltage 8 $\mu$V at a frequency 17 Hz. For the measurement of the impact of finite resistance, we added external resistors at room temperature to the symmetric junction  circuit. 

\section{Effect of asymmetric  tunneling resistance }

In this section we consider solely  the effect of the asymmetric tunneling resistances on the SINIS device characteristics.  We assume  $ R_{N} $ =0  such that $ R_{tot} $ = $R_{T}= R_{L}+R_{R}$.
First, we neglect the thermal effects and try to understand the results in a simple framework. Later on, we include the thermal effects. 

\subsection{Current-voltage and conductance characteristics} 

\begin{figure}
\includegraphics[width=0.95\columnwidth]{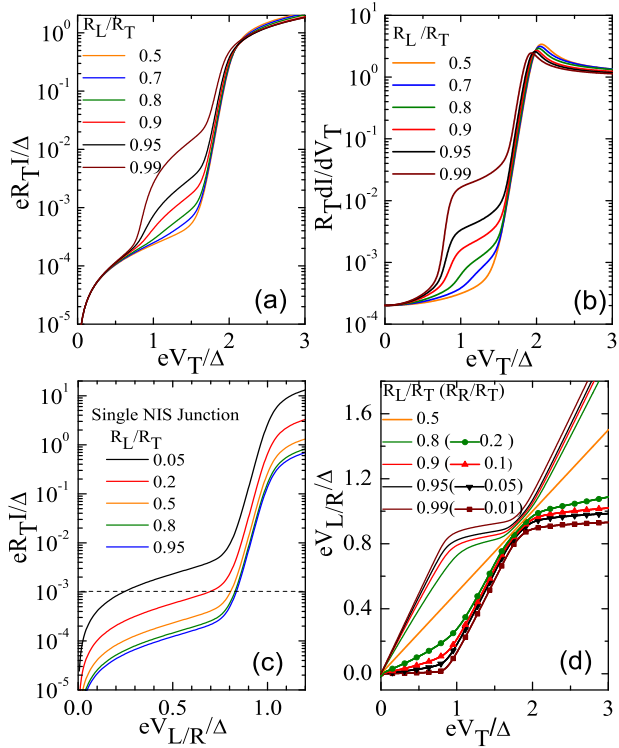}
\caption{[Color online] Calculated (a) current-total voltage and (b) differential conductance characteristics  of a SINIS device at  $T /T_{C}$ = 0.05 for different  level of asymmetry,  calculated in the framework of the electrical model.  Different  plots corresponds to different   combinations of $R_{L}$  and $R_{R}$ such that  the total tunneling resistance is constant. 
(c) Current-voltage characteristics of a single NIS junction at  $T /T_{C}$ = 0.05 for different  values of the tunnelling resistance. Dashed line gives $IeR_{T}/\Delta = \times$10$^{-3}$.  (d) Variation of the voltage of the individual NIS junctions $V_{L}$ and $V_{R}$  corresponding to some of the plots in (a) as a function of total voltage $V_{T}$.  }
 \label{fig3}
\end{figure}

In Fig.$~\ref{fig3}$ (a)  we show the evolution of the theoretical current - total voltage ($I-V_{T}$) characteristics of a SINIS tunnel junction pair with an increasing amount  of asymmetry in the  tunneling resistances, calculated for  $T_{S}/T_{C}$ =0.05.  This calculation was carried out in the framework of electrical model assuming $T_{N}$ =$T_{S}$, and keeping the total tunneling resistance $R_{T}=R_{L}$ + $R_{R}$ constant, but varying the values of the individual tunneling resistances.  The case where  $R_{L} = R_{R}$ = 0.5 $R_{T}$  is the   symmetric case, and only values of $R_{L} \geq 0.5 R_{T} $ are plotted because the results are symmetric with respect to the interchange of $L$ and $R$ junctions. It is clearly seen that any deviation from the symmetric case gives rise to an excess current in the intermediate subgap  region ($eV_{T}=\Delta .. 2\Delta$), whose magnitude increases with the asymmetry. Above the subgap region the current is reduced with increasing asymmetry. The magnitude of the current at $eV  = 1.5\Delta$  for $R_{L}/R_{T}$ =  0.95 is about one order of magnitude higher than in the symmetric case, while  the increase in current for a smaller asymmetry of  $R_{L}/R_{T}$ =  0.7  is just few tens of percent. The differential conductance $dI/dV_{T}$ curves corresponding to Fig.$~\ref{fig3}$(a)  are shown in Figure  $~\ref{fig3}$(b).  We see that a large enough asymmetry gives rise to an apparent secondary quasiparticle DOS "shoulder" between $\Delta ... 2\Delta$,  with an concomitant downward shift of the $\approx 2\Delta$ DOS peak and a depletion beyond $2\Delta$.   However,  far beyond the gap-edge,  all the current-voltage and conductance curves  merge. 

Let us now try to understand the origin of the excess current  in the framework of circuit theory. Fig.$ \ref{fig3}$(c)  shows the $I$-$V$  characteristics of a single NIS junction for different values of the tunnelling resistance at $ T_{S}/T_{C} =0.05 $. Clearly,  the shape of the curves are exactly the same, but simply scaled by the value of the tunnelling resistance, as seen from. Eq. (\ref{I}). By considering a constant current bias in the region where one of the junctions is near the gap edge, whereas the other not (for example, the dashed line), we can see that the total voltage drop $V_{T}=V_{L}+V_{R}$ is smaller for higher asymmetry, or,  conversely, for a constant total voltage the current is higher,  explaining thus the origin of  the excess subgap current.  To understand better why the excess current is pronounced around $V_{T}  = \Delta ..2\Delta$, one can study how the individual voltages $ V_{L}$ and $V_{R}$ vary  as a function of the total voltage $V_{T}$ across the SINIS junction, shown in   Fig.$ \ref{fig3}$(d). For a low total voltage $V_{T} < \Delta$, both junctions are in the subgap region where the junction resistances $R_{Ji}=V_i/I$ are much higher than the  tunnelling resistances $R_i$, but their ratio is the same as the ratio of the tunneling resistances $R_{L}/R_{T}$. Therefore,   most of the voltage drops across the junction with the larger resistance ($R_{L}$ in Fig.$ \ref{fig3}$(d)). With $V_{T} > \Delta$, however, the resistance of this junction drops dramatically below that of the other junction as the gap edge has been reached, thereby  giving rise to a increase in the current, saturation of the junction voltage $V_{L}$, and a much faster increase of the voltage of the other junction, which has now become the high resistance junction (still in subgap).   Finally, the second junction reaches the gap edge  ($V_{T} \geq 2\Delta$), and the roles of the junctions are again switched back to the original,  with voltage drops increasing as if the junctions are simple resistors with zero bias offsets $V_{L}=V_{R} \sim \Delta$. 

\begin{figure}
\includegraphics[width=1\columnwidth]{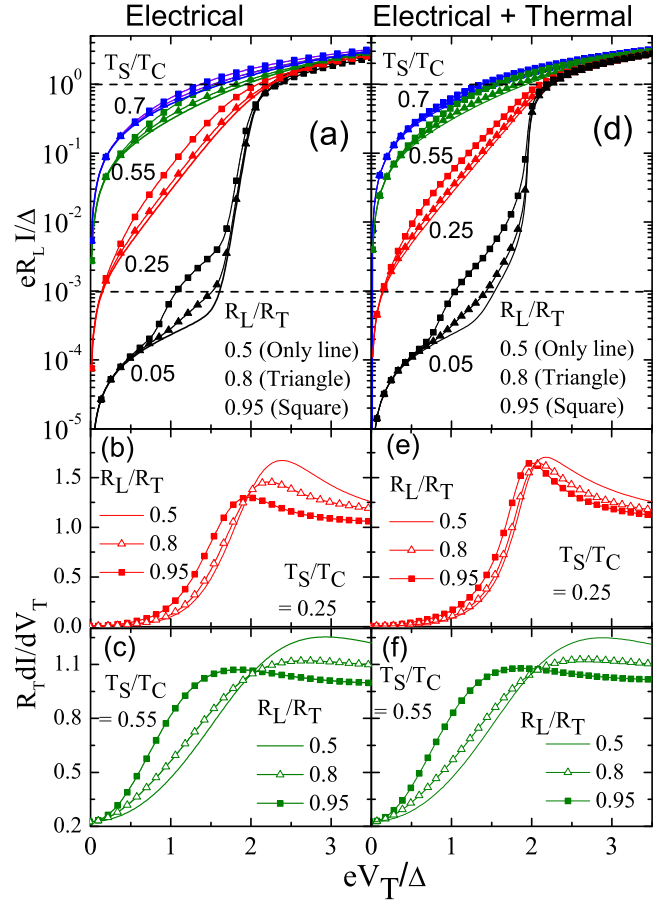}
\caption{[Color online] (a) Calculated current-voltage  characteristics  of a SINIS device at four bath temperatures ($T/T_{C}=$ 0.05,0.25, 0.55 and 0.7), and for three different amounts of asymmetry $ R_{L}/R_{T}$ = 0.95, 0.8 and 0.5, calculated without taking in account thermal effects. The corresponding conductance characteristics  calculated at bath temperatures of  $ T_{S}/T_{C} $  = 0.25 and  0.55 are shown in (b) and (c), respectively. (d), (e), (f) are the same, but now including the effect of the thermal circuit as well, with $B'=200$. 
} \label{fig4}
\end{figure}

Next, we consider the evolution of the electrical characteristics   of  an asymmetric SINIS device with the bath temperature.  In Fig.  $ \ref{fig4}$  (a)  we show the  $I$-$V_{T}$  characteristics  with three different asymmetry levels at four distinct bath temperatures, whereas in Fig.  $ \ref{fig4}$  (b) and (c) we plot $dI/dV_{T}$ at two different bath temperatures. Clearly, the effects of asymmetry   are pronounced in the entire temperature regime below $T_C$.   The distinct bump in the   current-voltage   and the conductance characteristics at $V_{T}  = \Delta ..2\Delta$  smears with increasing temperature into a broader feature, resembling a shift in $V$ at intermediate temperature range (e.g. at $ T_{S}/T_{C} = 0.25$). From the conductance curves we observe that  the height of the conductance maximum (at $V_{T} \approx  2\Delta$ for low asymmetry) decreases and shifts to lower values of $ V_{T}$ as the asymmetry increases. The higher the temperature, the larger this shift is: for example at $ T_{S}/T_{C} = 0.55$ and for $ R_{L}/R_{T}$ = 0.95  the relative shift is $\sim 40$ \%. The observed effects thus have the implication that if one does not know the level of asymmetry, an error in the determination of $\Delta$ from the data can follow. Figs. \ref{fig4}(d)-(f) show the same results, but now including the thermal circuit, Eq. \ref{th}. The general effect of asymmetry is the same, with only slight modifications to $I$-$V$ and $dI/dV$-$V$ characteristics in comparison to the pure electric model, due to self-heating and cooling effects.       

\begin{figure}
\includegraphics[width=0.95\columnwidth]{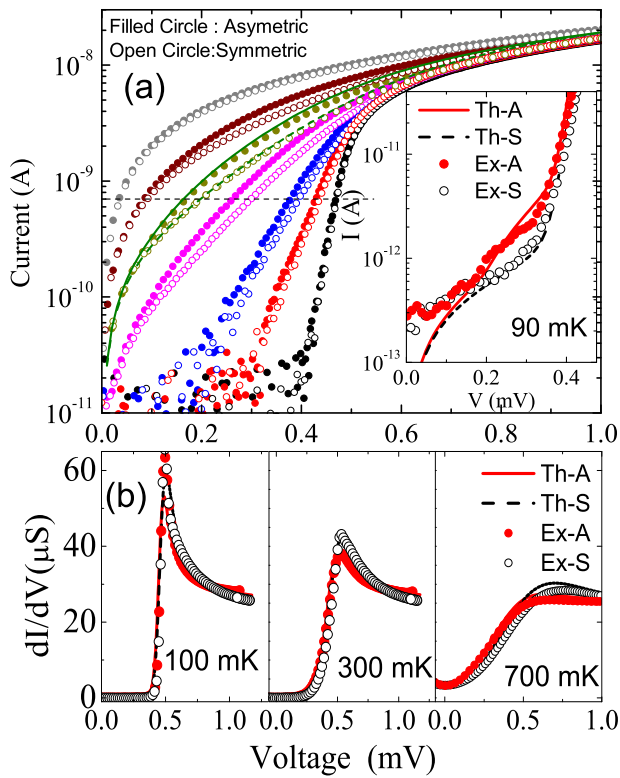}

\caption{[Color online] (a)Measured current-voltage characteristics  of the SINIS tunnel junction device of Fig. \ref{sem} at bath temperatures $T=$0.09,0.2,0.3,0.5,0.7,0.9 and 4.2 K, for the symmetric (open circles), and asymmetric (filled circles) junction pairs with $ R_{L}/R_{T} $ $\sim$ 0.875. The lines at 0.7 K (dashed for symmetric, solid for asymmetric)  are the theory curves  calculated using parameters corresponding to the sample.   
The inset  shows the results of a measurement at 90 mK in the deep subgap region, including the theory curves (solid for asymmetric, dashed for symmetric). 
(b) The  measured differential conductance characteristics at bath temperatures $T=0.1$, 0.3, and 0.7 K for both the asymmetric pair (filled circles) and symmetric pair (open circles).  The solid and dashed lines are the theoretical curves for the asymmetric, and symmetric junction pairs, respectively.} 
 \label{fig5}
\end{figure}

 In order to verify our computational results, we carried out experiments to compare the characteristics of an asymmetric junction to its symmetric counterpart with the sample of Fig. \ref{sem}. The results are shown in Figure  $\ref{fig5}$ (a) for the current-voltage, and (b) for the conductance characteristics \cite{jpcp}. In general, the experimental data resembles the theoretical plots of Fig.  \ref{fig4}  (a) quite well.  The slight 4 \% offset in current between the two curves at 4.2 K (where junctions are normal) is simply due to the 4 \% difference in the total tunneling resistances $R_{T}$. In agreement with the numerical results, there is clearly a voltage dependent excess current in the asymmetric device at temperatures below 4.2 K (for example an excess of 40 \% at 0.3 mV at 500 mK). At the lowest measured bath temperature of 90 mK, the predicted hump structure in the asymmetric device is clearly visible, shown in the inset of Fig.  \ref{fig4}  (a).   For comparison with theory, we have also plotted theoretical curves without free parameters at  90 mK (inset) and at 700 mK, with the level of asymmetry determined from the SEM image (difference in junction areas). The match with experiment is very good. 
   The  conductance characteristics measured at three distinct bath temperatures are also shown in  Fig.$\ref{fig5}$ (b), including the theory curves. The predicted lowering and shift of the conductance peak is visible in the experimental data, especially at the higher temperatures. The theoretical curves also match quite well.  
  
\subsection{Cooling power characteristics}
 \begin{figure}
\includegraphics[width=1\columnwidth]{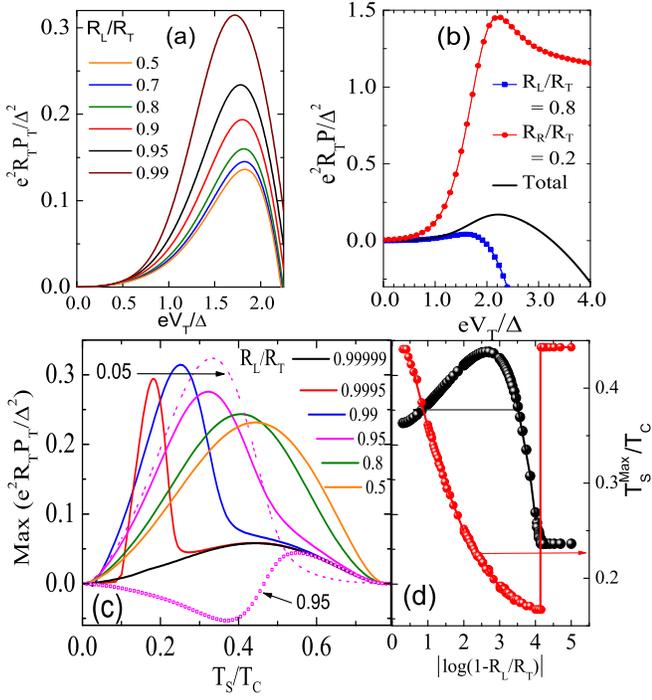}
\caption{[Color online] (a)Variation of  the total cooling power of the normal metal of asymmetric SINIS junctions as a function of applied voltage at $T_{S }/T_{C}$ =0.25, with varying asymmetry.  (b) Voltage dependence of cooling power of the individual $ L $ and  $ R $ junctions when $R_{L}/R_{T}=0.8$ (lines with symbols), together with the total cooling power (line). $T_{S }/T_{C}$ =0.25. (c) Variation of the maximum  cooling power with temperature and asymmetry. The peak shifts to lower temperature with increasing asymmetry. The dashed line and the hollow symbols correspond to individual $ L $ and $ R $ junctions when $ R_{L}/R_{T}$ = 0.95. (d) The variation of the maximum cooling power (left axis, black symbols) and the temperature of the peak in (c) (right axis, red symbols) vs. logarithmic deviation from $R_{L}/R_{T}=1$.} 
\label{fig6}
\end{figure}

 The next interesting question is, how are SINIS device cooling characteristics affected by the asymmetry? Figure  $~\ref{fig6}$(a)   shows the calculated  bias voltage dependence of the total equilibrium cooling  power of a SINIS device at $T_{S}/T_{C}$ =0.25 (where cooling is strongest) for different levels of asymmetry, but keeping $R_{T}$ constant. We see that the cooling power is actually {\em smallest} for the symmetric junction case,  and increases with asymmetry. For the highest asymmetry plotted, the enhancement is quite significant, a factor of 2.4  at the maximal cooling power bias point $ V_{T} \sim 1.7-1.8 \Delta $. Fig. \ref{fig6} (b) shows how the cooling power is divided between the two junctions vs. $V_{T}$ in the case where  $R_{L}/R_{T}=  0.8$.  Interestingly, the individual junctions exhibit  peak cooling powers at different values of the bias voltage, as the individual junctions reach the gap edge at different times. However, the total cooling power is still singly peaked, even though there are bias values and bath temperatures, where only one junction cools and the other one heats the normal metal.  
 
In order to investigate the cooling aspects further, in Fig.$~\ref{fig6}$(c)  we show the variation of the maximum cooling power  with bath temperature for various levels of asymmetry.  These calculations were carried out taking into account the temperature dependence of the BCS gap, $\Delta(T)$.  What we see is that at low  temperatures ($ T_{S}/T_{C} < 0.4-0.5$)  the peak cooling power increases with increasing asymmetry, while at higher temperatures the peak cooling power drops with increasing asymmetry. The optimal temperature value for cooling [maximum of the peak in Fig. \ref{fig6}(c)] shifts to lower temperatures, and the operational range in temperature (width of the peak) narrows  with increasing asymmetry. The optimal cooling power value keeps increasing up to a very large asymmetry $R_{L}/R_{T} = 0.999$, after which the enhancement is quickly destroyed, so that in the limit $R_{L}/R_{T} \rightarrow 1$, the single junction cooling characteristics are restored [Fig. \ref{fig6} (d)].  Fig.$~\ref{fig6}$(c) also shows the variation of the cooling power in the individual $ L $ and  $ R $  junctions for the case where $ R_{L}/R_{T}$ = 0.95.  It is seen clearly the that junction with lower resistance has much higher cooling power, but  the high resistance junction compensates by heating below $T_{S}/T_{C} = 0.5$.

\subsection{Impact of thermal effects: Cooling and negative differential resistance }
 \begin{figure}[htbp]
\includegraphics[width=0.93\columnwidth]{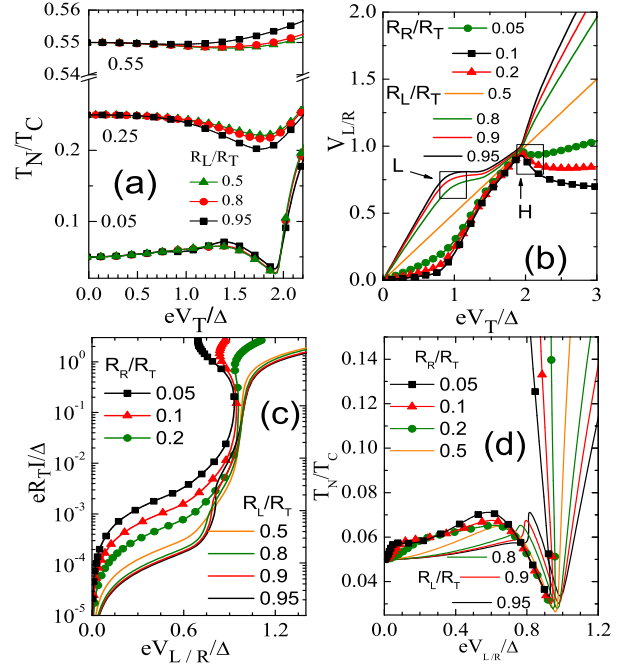}
\caption{[Color online] (a)  The variation of temperature of the normal metal $T_{N}$ as a function of the bias voltage and asymmetry at different bath temperatures $ T_{S}/T_{C} $  =0.05, 0.25, and 0.55.  (b) The variation of the voltage drop across the individual high- ($L$, lines) and low-resistance ($R$, lines with symbols) junctions as a function of the total voltage, calculated  at  $ T_{S} $  =0.05 and taking in account the thermal circuit,  for three different amounts of asymmetry. The markers $ L $ and $ H $ indicate the regions where negative differential resistance is observed.   (c) Current -  voltage characteristics of the individual $L$ (lines) and $R$ (lines with symbols) junctions at  $ T_{S}/T_{C} $  =0.05,  corresponding to the the plots in (b).  (d) The variation of $T_{N}$ as a function of the voltage  drop across  the individual $L$ (lines) and $R$ (lines with symbols) junctions at $ T_{S}/T_{C} $  =0.05. } 
\label{fig7}
\end{figure}

A more accurate picture of the electrical  characteristics can be obtained by taking into account the thermal circuit, Eq. \ref{th}, as well. This now enables us to calculate the electron temperature of the normal metal $T_{N}$  as a function of the applied bias voltage. Initially, we do not include the Joule heating in the normal metal and set $ R_{N} = 0$, later in section \ref{rn} we also consider the finite normal metal resistance.    The current-voltage and the conductance  curves were already shown in  Fig.$~\ref{fig4}$ (d)-(f). The main influences of the thermal circuit are that  the excess current feature spreads out to a bit higher voltages, and the conductance peak shift and reduction is not quite as prominent, especially around $T_S$ where cooling is strong [Fig. \ref{fig4} (e)].   

For the cooling characteristics, the variation of $ T_{N} $   with bias voltage and three different asymmetries are shown in Fig.$~\ref{fig7}$ (a)  for $ T_{S}/T_{C} $ =0.05, 0.25 and 0.55, using the scaled e-p parameter $B'=200$.  At $ T_{S}/T_{C} $  =0.05, the electron temperature first increases with increasing bias voltage, because of the finite Dynes parameter heating.  This heating gives way to the usual strong cooling  in the vicinity of the gap edge,  with $ T_{N} $ reaching its minimum at $V_{T} \approx 2 \Delta$.  Curiously,  although the cooling power  increases with increasing asymmetry (Fig.$~\ref{fig6}$), $T_{N}$ is not much affected.   This apparent contradiction appears because the cooling power also depends directly on $T_{N}$, and the asymmetry enhancement effect disappears for $T_{N} << T_{S}$. At the intermediate range $ T_{S}/T_{C} $  =0.25, however, the enhanced cooling power is clearly seen as temperature reduction at $V_{T}=1.7 \Delta$. On the other hand, at the highest bath temperatures plotted, asymmetry leads to a clear increase of $T_{N}$ due to the loss of cooling power at higher values of $T_S$, as seen in Fig. \ref{fig6} (c). 

 In Fig.$\ref{fig7}$ (b) we show the variation of the voltages  $ V_{L}$ and $V_{R}$  as  a function of total applied bias voltage across the SINIS device at $T_{S}/T_{C}=0.05$, including the thermal circuit. By comparing Fig.  $~\ref{fig7}$(b) to Fig.  $~\ref{fig3}$(d), we see that the most distinct feature that appears with the thermal effects in high asymmetry devices  is the abrupt {\em decrease} and a subsequent minimum in the voltage across the low resistance junction ($R_R$), above $V_{T} = 2\Delta$. This region is   marked by a box and label H for better visibility.  In order to understand the origin of such a phenomenon,  one has to look at the I-V characteristics of the  corresponding  individual junctions, shown in Fig. \ref{fig7}(c).   Clearly,  the junction with the lower tunneling resistance ($R_{R}$) exhibits negative differential resistance (NDR)  at  the high current range,  while the corresponding higher resistance junction exhibits a much weaker NDR at a lower current region around $IeR_{T}/\Delta = 10^{-3}$.  Junctions with asymmetry $R_{L}/R_{T} 0.5$- 0.6 (not shown here) exhibit no such NDR effect, nor do we observe any NDR in the pure electrical model, so that the effect is thermally driven. By observing the variation of $ T_{N} $  as a function of $V_{L}$ and $V_{R}$ (shown in Fig.$ \ref{fig7}$ (d)), we notice that NDR occurs at regions where there is a sharp change of $ T_{N} $.  One should also point out that the observed NDR effect is completely stable: the total I-V characteristics never exhibit the NDR effect, as one junction always stabilizes the other.  Also, the NDR effect occurs only in the low temperature region, surviving up to $T_{S}/T_{C} \sim 0.3$ for asymmetry $R_{L}/R_{T}=0.95$. A more detailed explanation of the NDR effect is given in the Appendix.
  
\begin{figure}[]
\includegraphics[width=1\columnwidth]{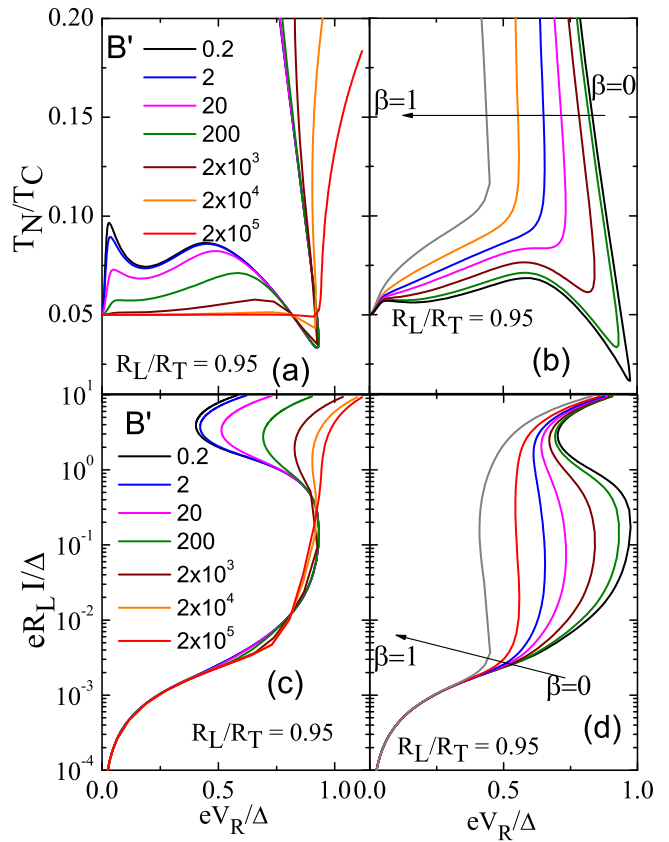}
\caption{[Color online](a) Variation of $T_{N}$ with $V_{R}$ for $ R_{L}/R_{T} $  =0.95, for (a) various values of parameter $B'$ at $ T_{S}/T_{C} $ =0.05 and $\beta=0.03$, and (b) for different values of $\beta$=0,0.03,0.1,0.2,0.3,0.5,1, with $B'=200$. (c) I-V characteristics of the individual low-resistance junction $ R $  corresponding to (a).  Strong signature of NDR is observed only for small values of $B'$. $ \beta $ = 0.03. (d) Variation of $T_{N}$  with $V_{R}$ for the values of $ \beta $ in (b), at $ T_{S}/T_{C} $ =0.05, $B'=200$.} 
\label{fig8}\end{figure}

 The fact that the strong switching in  $ T_{N} $  gives rise to the NDR has been investigated further by considering the variation of  $ T_{N} $  as a function of the applied bias for different values of the parameter $B' \propto BR_{T}$, which quantifies the ratio of heating or cooling by the substrate (e-p interaction) to the heat flow through the junctions. In practice, this could be changed by changing the normal metal volume.  In figure  $\ref{fig8}$(a)  we show the variation of  $ T_{N}$ as a function of voltage drop across the lower resistance junction $V_{R}$, when $R_{L}/R_{T}=0.95$, for various values of $B'$.  The corresponding current-voltage characteristics of the individual NIS junction are shown in $\ref{fig8}$(c).  Clearly, the NDR effect is strong for small values of $B'$ and vanishes  when  $B'> 10^{5}$. Similarly, when the backflow of heat from the superconductor is increased (parameter  $ \beta $), the  dip in the value of $ T_{N} $ decreases (less cooling), resulting in the reduction of the rapid temperature change, and the NDR effect, as seen in Figs. $\ref{fig8}$(b) and (d).  For a different level of asymmetry, the parameter region in $(B',\beta)$ plane where the NDR occurs, changes.

\subsection{Thermometry}
\begin{figure}[]
\includegraphics[width=1\columnwidth]{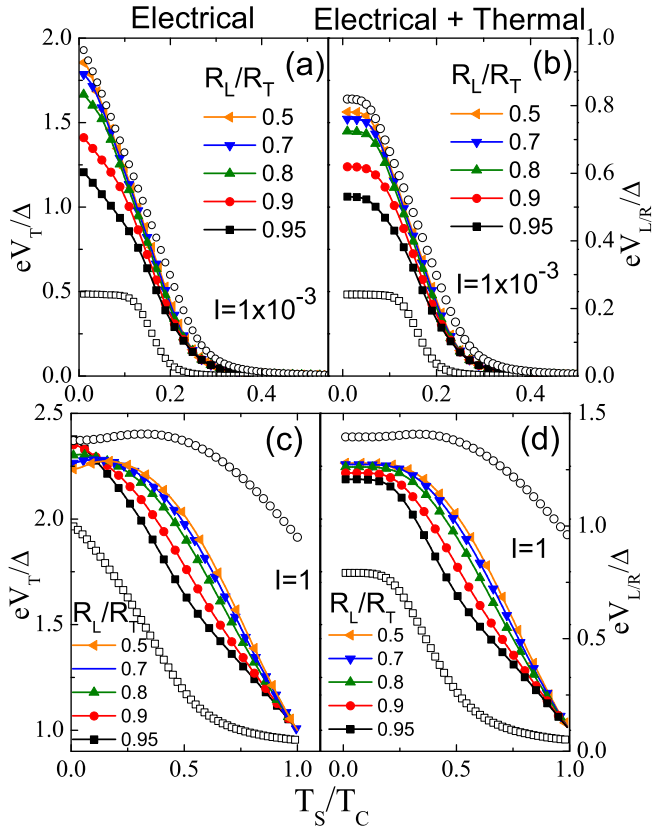}
\caption{[Color online]Variation of the total SINIS junction voltage, $V_{T}$ (lines-symbols), for different values of asymmetry, as a function of bath temperature $T_{S}$, for bias currents (a) $IeR_{T}/\Delta 1\times$ 10$^{-3}$, and (c) $IeR_{T}/\Delta=$1 in the  electrical model. With thermal effects taken into account, (a) and (c) are modified to (b) and (d), where we used $B'=200$.  The open symbols show the individual junction voltages $V_{L}$ and $V_{R}$ corresponding to  $R _{L}/R_{T}$ =0.95 (right axis scales). } \label{fig11}
\end{figure}
\begin{figure}[]
\includegraphics[width=0.8\columnwidth]{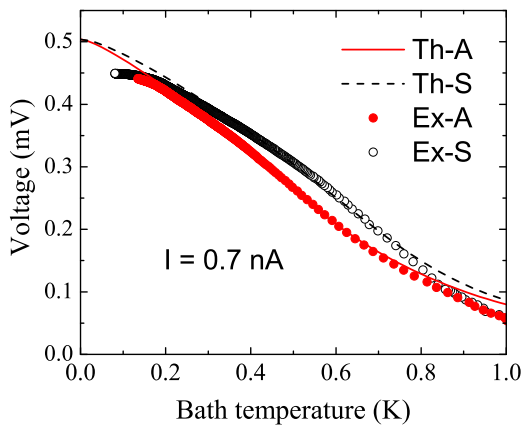}
\caption{[Color online]Measured $V_{T}$  vs.  bath temperature for both the symmetric (open circels) and asymmetric (filled circles) junction pair of the sample, with $ R_{L}/R_{T} $ $\sim$ 0.875 for the asymmetric case. The current bias used was  0.7 nA.  The lines show the theorical curves calculated using known parameters of the sample obtained from I-V curves.} \label{figexpth}
\end{figure}

For thermometry, a SINIS device is usually current biased, so that the thermometric performance can be characterized by plotting the $V_{T}$ vs. $T_{S}$ dependence. This is shown for an asymmetric  SINIS device in Fig.$ \ref{fig11}$, calculated using both the purely electrical (Figs.$ \ref{fig11}$ (a) and(c)) and electrical+thermal models (Figs.$ \ref{fig11}$ (b) and(d)),   for two different values of bias current.  Interestingly, for the low bias current ($ IeR_{T}/\Delta = $ 1$\times$ 10$^{-3}$), the symmetric thermometer has the best responsivity,  $dV/dT$.  In addition, if the thermal circuit (self-heating) is taken into account,  the  thermometer saturates below  $T/T_{C}\sim $ 0.05 for all values of asymmetry (Fig. \ref{fig11} (b)). This saturation is due to the electron-phonon decoupling; $T_{N}$ does not follow $T_{S}$ anymore, and $V_{T}$ saturates to a value that is determined by the self-heating effects (value of $B'$).  Thus, for thermometric applications,  it is desirable to minimize the effects of the thermal circuit by maximizing $B'$, which can be achieved by increasing the normal metal volume, or by increasing $R_{T}$. At a bias current high enough beyond the subgap regime $IeR_{T}/\Delta = $ 1, shown in Figs. Fig. \ref{fig11} (c) and (d),  an asymmetric  junction device has a better responsivity than the symmetric one  at low temperatures $T/T_{C} < 0.4$, where the symmetric device response dies out. However, with the thermal circuit, the response  again saturates even for the asymmetric junctions below    $T/T_{C} \sim$  0.25, leaving only a smaller window of improved performance around $0.25 <  T/T_{C} < 0.4$ for the value of $B'=200$ used here. Figures \ref{fig11} (a)-(d) also show the variation of the voltages across the individual junctions,  showing that the performance degradation  at low bias currents is caused by the saturation of the low resistance junction, whereas at high bias the low resistance junction gives the performance boost discussed above. In conclusion, for high bias currents, a highly asymmetric junction can act as a good  thermometer over a larger temperature range, especially if the thermal circuit effects are minimized. This is because  one of the junctions acts as a good thermometer at low temperatures, while the other one has better responsivity at higher temperatures.

Finally, we have experimentally checked the effects of asymmetry on thermometric performance for the device discussed in section \ref{sample}.  Fig. \ref{figexpth} shows the measured $V_{T}$ vs $T_{S}$ curves for both the asymmetric and symmetric junction pairs, for a bias current of 0.7 nA corresponding to $IeR_{T}/\Delta \sim 0.13$, together with the corresponding theory curves of the electrical model. The symmetric theory has no fitting parameters, whereas the asymmetric theory has only one, the level of asymmetry $R_{L}/R_{T} \sim$ 0.875. This corresponds well with the ratio of junction areas from the SEM image (0.83). It is clear that the general trends of the changes due to asymmetry are reproduced, including the improvement in responsivity below $T < 0.6$ K. The discrepancy between theory and experiment at high temperatures is not fully understood at the moment.

\section{Effect of finite series  resistance}
\label{rn} 
In this section, we consider the effect of finite series resistance on the conductance, cooling and thermometry aspects of a SINIS device.
As already mentioned earlier, the finite series resistance always models the resistance of the normal metal island $R_{N}$, but could also include line and contact resistances, if the measurement is performed in two-probe configuration. In the thermal circuit, only $R_{N}$ appears in the Joule heating term in Eq. \ref{th}, as the Joule heat of the other series resistors is always dissipated far from the active device area.  For the sake of simplicity, we consider here the case of a symmetric SINIS device only,  $ R_L $ = $ R_R $, with results shown for both the electrical and electrical + thermal model.  Furthermore, for the electrical + thermal model, we only include the most important resistance $R_{N}$, and do not consider the  effects of other series resistances,  which could be easily included in  the calculations by adding them in the electrical circuit equations. 

\subsection{Impact on conductance characteristics}

\begin{figure}[]
\includegraphics[width=1\columnwidth]{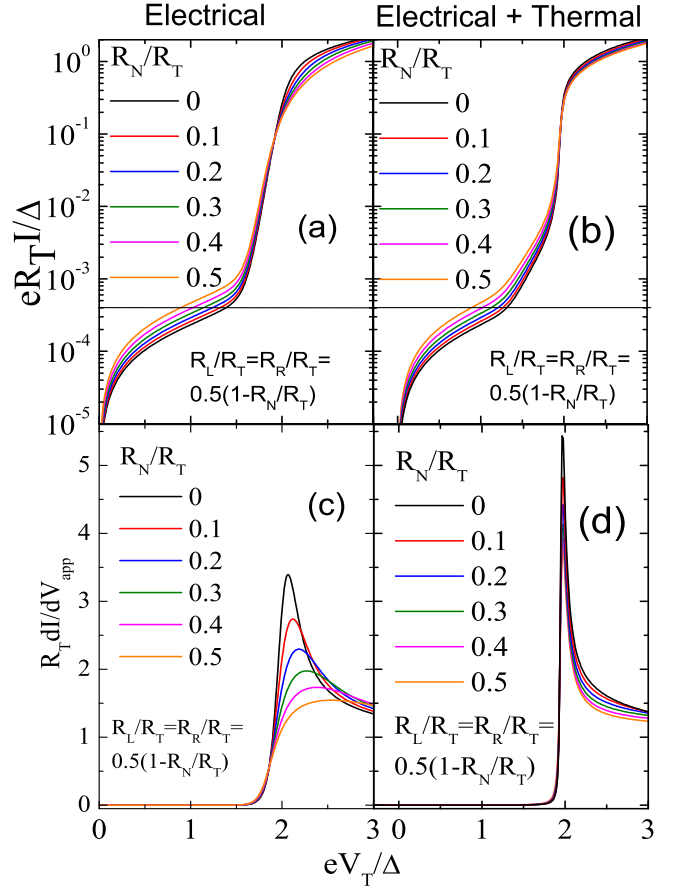}
\caption{[Color online] Calculated I-V characteristics of a symmetric SINIS device vs. $R_{N}$,  when the total resistance $R_{T} $= $ R_{L}+ R_{R}+ R_{N}$  is constant in the (a) electrical and (b) electrical + thermal case.  $T_{S}/T_{C}$ = 0.05.   The conductance characteristic corresponding to (a) and (b) are displayed in (c) and (d). } \label{fig12}
\end{figure}
While discussing the effect of finite normal metal resistance on SINIS characteristics we consider two different cases:

(i) In the first case,  the value of  $ R_{N}$ is varied keeping the total resistance constant (i.e. $R_{T}$= $ R_{L}+ R_{R}+ R_{N}$ =const.).  This situation is typical in actual experiments, where  the value of  $ R_{N}$  remains unknown but $R_{T} $ is known precisely. In Fig.$ \ref{fig12}$  (a) and (c) we show the current-voltage and conductance characteristics as a function of $R_{N}$, calculated at  $T_{S}/T_{C}$ = 0.05 for the electrical model, and in Fig.$ \ref{fig12}$  (b) and (d) for the electrical + thermal model, respectively.   It is  clearly seen that in both models, the subgap current actually increases with increasing magnitude of  $R_{N}$, whereas above the gap the current decreases with increasing $R_{N}$, until finally all curves converge asymptotically at high voltages  (not shown). This initially counterintuitive behavior can be explained as follows: In the subgap region, the resistances of the junctions are  much larger than values of $R_{N}$ considered here, so that the current is limited by the junctions. In that case, increase of $R_{N}$ corresponds to a decrease of $R_{L}$+$R_{R}$, which leads to increase in the current. However, above the gap the junction resistances become comparable to $R_{N}$ values considered, and voltage starts to drop across $R_{N}$. In that case, an increase of $R_{N}$ leads to a decrease of the voltage across the junctions, leading to a decrease in current. In terms of differential conductance, the main effect in the electrical model (Fig. \ref{fig12} (c)) is a decrease of the conductance peak value, with a concomitant flattening and an {\em outward} shift of the peak position.  If one considers the thermal circuit,  the flattening of the conductance peak can be arrested as a result of the self-cooling and the Joule heating in $R_{N}$, but the conductance maxima still drops with increasing $ R_{N}$, as seen from Fig.$ \ref{fig12}$(d).  Thus, similar to the case of asymmetric junctions, the net effect of finite $ R_{N}$ is an excess current in the subgap and a reduction of current above the gap. However, the conductance curves behave in a different way both in the subgap and in the conductance peak region, as seen by comparing Figs. \ref{fig4}(b) and \ref{fig12}.
\begin{figure}[]
\includegraphics[width=1\columnwidth]{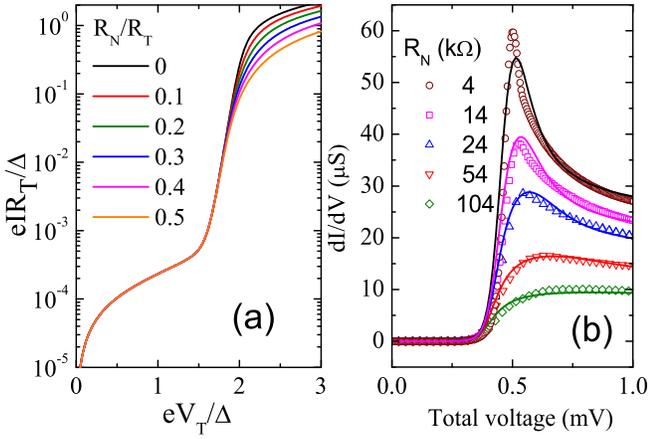}
\caption{[Color online] (a) Calculated I-V characteristics of a symmetric SINIS device vs. $R_{N}$, when $R_{N}$ varies independently of $R_{L} +  R_{R}$, for the electrical case. $T_{S}/T_{C}$ = 0.05. (b) Experimental data (symbols) for conductance curves with varying external series resistor (from 4 k$\Omega$ to 104 k$\Omega$) measured at $T=0.15$ K , together with the theoretical characteristics (lines) corresponding to the experiment.} \label{fig13}
\end{figure} 

(ii) In the second case, $ R_{N}$ is assumed to vary independently of $R_{L} + R_{R}$, corresponding experimentally to the case where different samples are compared. In Fig.$\ref{fig13}$  (a)  we show the current-voltage  characteristics calculated at $T_{S}/T_{C}$ = 0.05   for the electrical model.  Now,  the subgap current is insensitive to the value of $R_{N}$, whereas the current above the gap is reduced due to the direct effect of $R_{N}$ limiting the current. The addition of the thermal circuit gives the same result (not shown), with only the shape of the I-V curve changing slightly [as seen in Fig. \ref{fig12} (b).]  The behavior of the conductance in the electrical model is also similar to case (i), with the only difference that the high voltage asymptotes naturally do no converge. This is seen in Fig. \ref{fig13} (b), where we also plot the experimentally measured characteristics, produced by changing an external series resistor. 
The data agrees well with the purely electrical model, clearly giving rise to a decrease and shift of the conductance maxima at the gap edge.

\subsection{Cooling} 
\begin{figure}[]
\includegraphics[width=1\columnwidth]{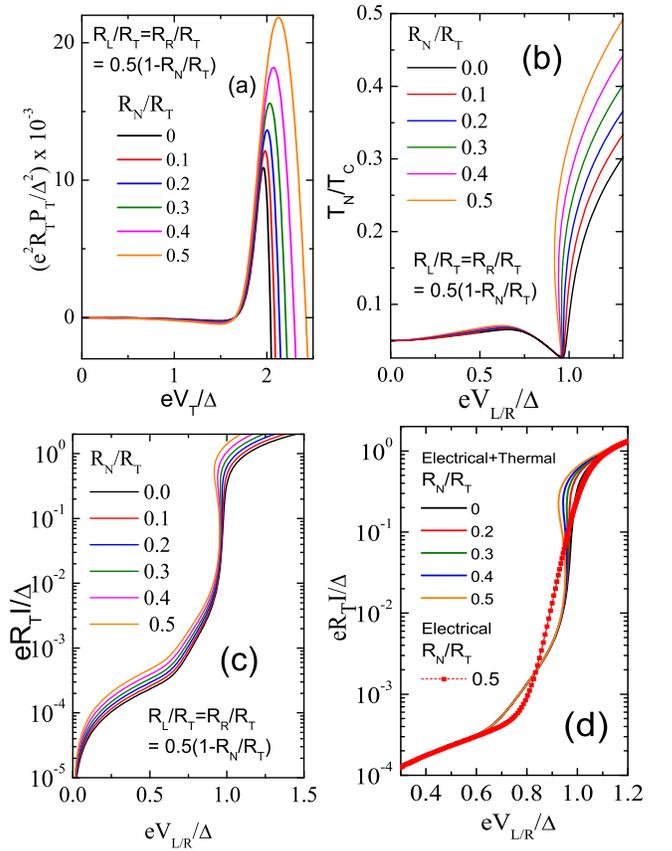}
\caption{[Color online] (a) Variation of the total cooling power of a symmetric SINIS device vs. $V_{T}$  for different values of $ R_{N}$ calculated at  $T_{S}/T_{C}$ = 0.05 with  total resistance   $R_{T}= R_{L}+ R_{R}+ R_{N}$ kept constant. (b)  $T_{N}$ vs individual junction voltage for different values of $R_{N}$, with $R_{T}$ constant. (c) I-V characteristics of the individual junction with varying $R_{N}$, for $R_{T}$ constant, taking in account the thermal circuit, and (d) the same but letting $R_{N}$ vary independently. The points also show the result in the purely electrical model for a high $R_{N}/R_{T}=0.5$.} \label{fig15}
\end{figure}
The variation of the total equilibrium cooling power ($T_{N}=T_{S}$) with total bias voltage $V_{T}=V_{L}+V_{R}+V_{N}$ and series resistance $R_{N}$ at   $T_{S}/T_{C}$ = 0.05  is shown in Fig.  $ \ref{fig15}$ (a) for the case where $R_{T}$ is kept constant.  We see that the peak cooling  power increases with  increasing $ R_{N}$, and shifts to a higher voltage. The power increase happens because $R_{L}+R_{R}$ has to decrease, thereby increasing the cooling power, whereas the shift is due to the finite voltage drop across $R_{N}$. In the case where $R_{N}$ varies independently, the cooling power naturally doesn't change at all, but curves simply shift due to the finite voltage drop across $R_{N}$ (not shown). In addition, if one investigates  the cooling  [Fig. \ref{fig15} (b)] and individual NIS junction I-V characteristics [Fig. \ref{fig15} (c)]  as a function of $R_{N}$ in the electrical+thermal model with $R_{T}$ constant, one finds again the NDR effect, whose origin is again associated with the strong increase of $ T_N $ as a function of voltage, with the back-bending corresponding to the NDR region. The minimum temperature is again not affected, similar to the asymmetric case [Fig. \ref{fig7} (a)].  Fig. \ref{fig15} (d) shows also the single junction I-V curves with $R_{N}$ varied independently, with one I-V curve calculated with the pure electrical model and a high value of $R_{N}= 0.5 R_{T}$ also shown. Clearly, no NDR appears in the electrical model. The NDR effect is once again fully stable, with the stabilization coming from the voltage drop across the resistance $R_{N}$.  
 \subsection{Thermometry} 
 \begin{figure}[]
\includegraphics[width=1\columnwidth]{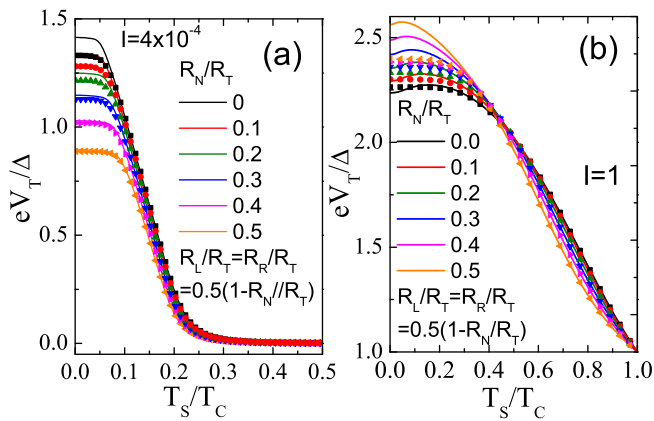}
\caption{[Color online] $V_{T}$ vs $T_{S}$ of a symmetric SINIS device for different values of $ R_{N}$, with a current bias $IeR_{T}/\Delta$ = 4$ \times$10$^{-4}$) in (a), and  $IeR_{T}/\Delta$ = 1 in (b). The total resistance   $R_{T}= R_{L}+ R_{R}+ R_{N}$ is kept constant. The lines show the characteristics in the purely electrical model, whereas the lines with symbols take into account the thermal circuit.}\label{fig16}
\end{figure}
The  thermometric characteristics, $V_{T}$ vs. $T_{S}$, for the first case where the $R_{T}$ is kept constant, are shown in  Figs. $\ref{fig16}$  (a) and (b) for two different bias currents $IeR_{T}/\Delta$ = 4$ \times$10$^{-4}$ (subgap) and $IeR_{T}/\Delta$ = 1 (above gap), respectively. Both the purely electrical and electrical+thermal results are shown. As already seen from  Fig. $\ref{fig12}$  (a), for a constant $I$ in the subgap region, the voltage drop across the SINIS pair decreases slightly as $ R_{N}$ increases. This is also seen in Fig. $\ref{fig16}$ (a), where it is clear that the effect on responsivity is small. However, the situation is reversed for the high bias current, Fig. $\ref{fig16}$ (b). Then the responsivity of the thermometer is clearly better for higher  $ R_{N}$ below $T_{S}/T_{C} = 0.4$, if only electrical circuit is considered. On the other hand, when $ R_{N}$  varies independently, no performance improvement or reduction is seen (not shown). At low temperatures the thermometers eventually saturate, with the saturation regime depending on the current bias value and the thermal circuit (overheating). This overheating can negate all the performance gains if $B'$ is not large enough, as seen in Fig. \ref{fig16} (b).

\section{Conclusions}

In this paper, we presented a comprehensive study of the effects of (i) asymmetry of the tunneling resistances and (ii) additional series resistance on the behavior of double junction SINIS tunnel junction devices.
Both experiments and calculations showed that  asymmetry  results in an excess current below the gap voltage and a reduction and shift of the conductance peak at the gap edge. This effect takes place regardless of the thermal circuit characteristics. In addition, our calculations predict that it is possible to observe negative differential resistance in the individual junctions in a realistic thermal environment, if asymmetry is high enough. 
Interestingly, we also calculate that the total equilibrium cooling power  of an asymmetric device can be larger than of the symmetric device with the same total tunneling resistance. This enhancement can be very large, over 200 \% for an extreme asymmetry of $R_{L}/R_{T}=0.99$. If one looks at the maximum cooling power as a function of the bath temperature, it shifts to lower temperatures with increasing asymmetry, and attains an enhancement of $\sim 30 $ \% over the symmetric device. The shift  of the cooling power maximum to lower temperatures is promising for real device operation at lower temperatures. However, the improved cooling power at equilibrium does not always translate into a lower minimum normal metal temperature. Finally, our theoretical results for thermometry show that in the low temperature regime $T/T_{C} < 0.2$, asymmetry gives typically no  benefits. However, at higher temperatures (using higher bias currents) one can take advantage of the different responsivities of the individual junctions in the asymmetric SINIS device, to construct a thermometer that is sensitive over a broader temperature range. 

We also showed experimentally and theoretically that a finite series resistance gives rise to a clear reduction of the conductance maximum at the gap edge, and that the thermal circuit affects the details significantly.  Our calculations show that it is also possible to observe the negative differential resistance effect in the individual junctions in a symmetric device with a high enough normal metal resistance.  For cooling power, a series resistance naturally offers no absolute benefits, however, with an unknown series resistance, the true cooling power is higher than what one would expect for the ideal case. In thermometry, there are also no benefits in the absolute sense, but an unknown series resistance means that the responsivity can be better than what one would expect for a current bias above the gap.    
\section*{Acknowledgements}
 This research has been supported by Academy of Finland project number 128532.

\section{Appendix} 

\begin{figure}[]
\includegraphics[width=0.75\columnwidth]{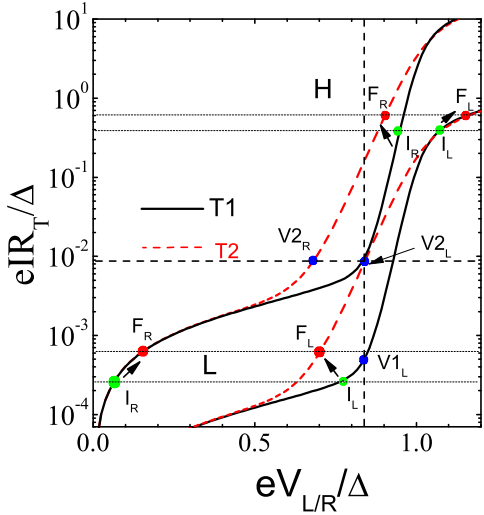}
\caption{[Color online]Schematic representation of the origin of NDR.  The initial state of the junctions $ L $ and  $ R $  at $ T_{1}$  are given  $ I_L $ and  $ I_R $ respectively and the corresponding final states by  $ F_L $ and  $ F_R $ respectively. When the increase in the bias voltage is sufficient to cause a  heating  of the normal metal from an initial temperature  $ T_{1}$ to a final temperature  $ T_{2}$, the junctions goes from  the initial state to final state via virtual states namely $ V1_L $   and $ V2_L $.  The initial, final and the virtual states are color coded as green, red and blue respectively. } 
\label{fig9}\end{figure}

In order  to understand the origin of the observed NDR due to asymmetry, let us consider  the specific case where  $ R_{L}/R_{T}$ = 0.95 and  $R_{R}/R_{T}$  =0.05. The   current-voltage characteristics of these individual junctions at two temperatures $ T_{1}$  and $ T_{2}$ ( where $ T_{2}$  $ > $ $ T_{1}$) are shown in Fig.$ \ref{fig9}$.  Let us consider the situation at a low bias current region (shown by marker $ L $). The initial state  of the junctions $ L $ and  $ R $  at $ T_{1}$  are given  $ I_L $ and  $ I_R $ respectively. At this point we increase the applied bias voltage by a small amount, which is sufficient to give rise to an increase of $ T_{N}$  from $ T_{1}$ to  $ T_{2}$.  Since the sub-gap resistance of junction $ L$ is  larger than that of junction $ R $, most of the increment in bias voltage drops across the $ L$ junction, thereby taking it from initial state $ I_L $ to a virtual state $ V1_L $.  We identify this state as virtual as we have not considered the heating yet. At this point we ''turn on'' the heating  which demands that the junction switches to point $ V2_L $ lying on the $ T_2 $ branch. Once  again, we identify  $ V2_L $  as a virtual state because if  junction $ L$   is in $ V2_L $,   junction $ R$  has to be in  $ V2_R  $ since they  are in series and  the same current flows through them. However, this is not possible, because in that case the sum of the voltage drops across the individual junctions exceeds the total applied voltage.  Thus, the only solution is that for the given temperature  $ T_{2}$ the total applied voltage is redistributed across the individual junctions in such a way that final configurations of the  $ L $ and  $ R $  are given  $ F_L $ and  $ F_R $ respectively.   As seen from the figure, this redistribution  gives rise to an increase in the current, increase in the voltage across $R$, a drop in the voltage across $ L $ junction, and thus leads to negative differential resistance (NDR) in the $L$ junction. Using similar arguments, the origin of NDR in the  $ R $ junction in the high current region ( corresponding to marker $ H $)  can be understood.

\end{document}